\documentclass[preprint,12pt]{elsarticle}

\usepackage{graphicx}

\usepackage{amssymb}

\journal{Planetary and Space Science}

\begin{document}

\begin{frontmatter}

\title{Observing the Earth as an exoplanet with LOUPE, 
       the Lunar Observatory for Unresolved Polarimetry of Earth}

\author[1,2]{Karalidi, T.}
\author[1]{Stam, D.M.}
\author[2]{Snik, F.}
\author[4]{Bagnulo, S.}
\author[3]{Sparks, W.B.}
\author[2]{Keller, C.U.}

\address[1]{SRON Netherlands Institute for Space Research, The Netherlands}
\address[2]{Leiden Observatory, The Netherlands}
\address[3]{Space Telescope Science Institute, USA}
\address[4]{Armagh Observatory, UK}

%-------------------------------------------------------------------------------------
\begin{abstract}
The detections of small, rocky exoplanets have surged in recent years
and will likely continue to do so. To know whether a rocky exoplanet
is habitable, we have to characterise its atmosphere and surface.  A
promising characterisation method for rocky exoplanets is direct
detection using spectropolarimetry.  This method will be based on
single pixel signals, because spatially resolving exoplanets is
impossible with current and near-future instruments.  Well-tested
retrieval algorithms are essential to interpret these single pixel
signals in terms of atmospheric composition, cloud and surface
coverage.  Observations of Earth itself provide the obvious benchmark
data for testing such algorithms. The observations should provide
signals that are integrated over the Earth's disk, that capture day
and night variations, and all phase angles.  The Moon is a unique
platform from where the Earth can be observed as an exoplanet,
undisturbed, all of the time.  Here, we present LOUPE, the Lunar
Observatory for Unresolved Polarimetry of Earth, a small and robust
spectropolarimeter to observe our Earth as an exoplanet.
\end{abstract}

%-------------------------------------------------------------------------------------

\end{frontmatter}

%-------------------------------------------------------------------------------------
\section{Introduction}
\label{Sec:Intro}

Since the discovery of the first exoplanet \citep{mayorqueloz95}, more
than 700 exoplanets have been detected as of today.  Even though most
of these exoplanets are gas giants, in recent years the number of
detected smaller mass planets has surged \citep[see
  e.g.][]{wordsworth11}.  Indeed, according to \citep{cassan12}, about
62\% of the Milky Way stars should have an Earth--like planet.  A
near-future detection of an Earth-sized exoplanet inside its star's
habitable zone seems unevitable.  Whether or not an Earth-sized planet
in a habitable zone is actually habitable, depends strongly on the
composition and structure of its atmosphere. As an example, the
Venusian surface is about 500$^\circ$ hotter than expected from Venus'
orbital distance and albedo, thanks to an extremely strong greenhouse
effect in its thick CO$_2$ atmosphere.  Hence, a characterisation of
the planetary atmosphere will be needed to address a planet's
habitability.

Currently, atmospheres of exoplanets are being characterised using the
so--called {\em transit method} \citep[see
  e.g.][]{beaulieu10,millerricci10}.  This method is based on
measurements of the wavelength dependendence of starlight that filters
through the upper layers of the planetary atmosphere during the
primary transit (when the planet passes in front of the star), or of
the planetary flux just before or after the secondary eclipse (during
which the planet passes behind its star). The transit method is mostly
applied to gaseous planets that orbit close to their star.
Earth-sized exoplanets in the habitable zone of their star are too
small to yield measurable transits \citep[][]{kaltenegger09}.

The best way to characterise the atmosphere and surface of an
Earth-sized exoplanet, is through {\em direct detection}, using large
ground--based telescopes such as the European Extremely Large
Telescope (E-ELT) or space telescopes with diameters of a few
meters. With direct detection the light of a planet is measured
separately from the stellar light (except for some background
starlight). But even if we observe an exoplanet with a direct
detection, the planet itself will be unresolved, i.e. it will appear
as a single pixel.  If the planet resembles the Earth, this single
pixel holds information on oceans and continents, coverage by
vegetation, desert, and, for example, snow and ice, all overlaid by
various types of patchy clouds.

Polarimetry promises to play an important role in exoplanet research
both for exoplanet detection and characterisation. In particular,
because the direct starlight is unpolarized, while the starlight that
is reflected by a planet will usually be polarized, polarimetry can
increase the planet--to--star contrast ratio by 3 to 4 orders of
magnitude \citep{keller10}, thus facilitating the detection of an
exoplanet that might otherwise be lost in the glare of its parent
star. Additionally, as in the case of Solar System planets \citep[see
  e.g.][]{hansenhovenier74,mishchenko97}, polarimetry will help the
characterisation of planetary surfaces and atmospheres, because the
polarisation of the reflected starlight is very sensitive to the
physical properties of an atmosphere and surface.  Combining flux with
polarimetric observations will also help to break retrieval
degeneracies that flux--only measurements have \citep[see
  e.g.][]{stam08, karalidi11, karalidi11b}. Finally, while measuring
the state of {\em linear} polarisation of reflected starlight helps to
characterise a planetary atmosphere and surface, the degree of {\em
  circular} polarisation of this light appears to be an indicator for
the existence of life on a planet, since circular polarisation, and in
particular its wavelength dependence, is linked to homochirality of
the complex molecules that are essential for life
\citep{sparks09,sterzik10}.

To decipher future signals of directly detected Earth-like exoplanets,
numerical models that can simulate single pixel signals of exoplanets
with inhomogeneous atmospheres and surfaces, are essential.  Such
models are essential for the design and optimisation of telescope
instruments and mission profiles (spectral bands, spectral resolution,
integration times, revisiting times, etc.), and, once observations are
available, they are a necessary tool to interpret the observations.
There are a number of numerical models that are used to calculate
signals of gaseous and terrestrial exoplanets, for reflected starlight
and/or thermally emitted radiation \citep[see e.g.][]{seager00,
  ford01, stam08, tinetti06c, karalidi11b}. In order to validate the
results of such numerical models, it is important to compare them
against observations.  The obvious test--case for numerical models for
Earth--like exoplanets, is Earth itself. To fully validate these
models, we need observations of the Earth as if it were an exoplanet,
hence single pixel observations that cover the diurnal rotations of
the Earth, and all phases of the Earth.  And, ideally, the
observations should cover different seasons to record the changes in
surface albedos and weather patterns.

An excellent location for performing such observations and for
building a benchmark dataset is the lunar surface facing the
Earth. From there, we can observe the whole disk of the Earth, all of
the time, at all phase angles, throughout the year. As we will argue
in more detail in Sect.~\ref{Sec:Moon}, such observations cannot be
achieved by e.g. combining observations of Low Earth Orbit (LEO)
satellites.  In this paper, we present LOUPE, the Lunar Observatory
for Unresolved Polarimetry of Earth. LOUPE is a small and robust
spectropolarimeter that measures the flux and state of polarization of
sunlight that is reflected by the Earth e.g. from ESA's Lunar Lander.

This paper is structured as follows. In Sect.~\ref{Sec:spectra}, we
present calculated flux and polarisation spectra of a single pixel
Earth.  In Sect.~\ref{Sec:Moon}, we summarize the advantages of
observing the Earth from the moon. In Sect.~\ref{Sec:LOUPE}, we
describe the LOUPE instrument.  Section~\ref{Sec:summary}, finally,
contains the summary and our conclusions.

%---------------------------------------------------------------------------------
\section{Flux and polarisation spectra of the Earth as an exoplanet}
\label{Sec:spectra}

%---------------------------------------------------------------------------------
\subsection{Flux and polarisation definitions}

Sunlight that is reflected by a planet can be described by a flux
vector $\pi\vec{F}= \pi [F, Q, U, V]$, with $\pi F$ the total flux,
$\pi Q$ and $\pi U$ the linearly polarised fluxes and $\pi V$ the
circularly polarised flux \citep[see e.g.][]{hansentravis74,
  hovenier04, stam08}. Each flux parameter depends on the wavelength
$\lambda$, and has dimensions W~m$^{-2}$m$^{-1}$. Linearly polarised
fluxes $\pi Q$ and $\pi U$ are defined with respect to the plane
through the center of the star, the planet and the observer \citep[see
  also][]{karalidi11b}.  The degree of polarisation $P$ of the
reflected sunlight is defined as the ratio of the polarised flux to
the total flux, thus $P= \sqrt{Q^2 + U^2 + V^2})/F$.  Specifically,
the degree of linear polarisation is defined as
%\begin{equation}
  $P_{\rm L}= \sqrt{Q^2 + U^2}/F$,
%\end{equation}
and the degree of circular polarisation as
%\begin{equation}
  $ P_{\rm C}= V/F$.
%\end{equation}

%------------------------------------------------------------------------
\subsection{Sample flux and linear polarisation signals of the Earth}

\begin{figure}
\centering 
\includegraphics[width=2.6in]{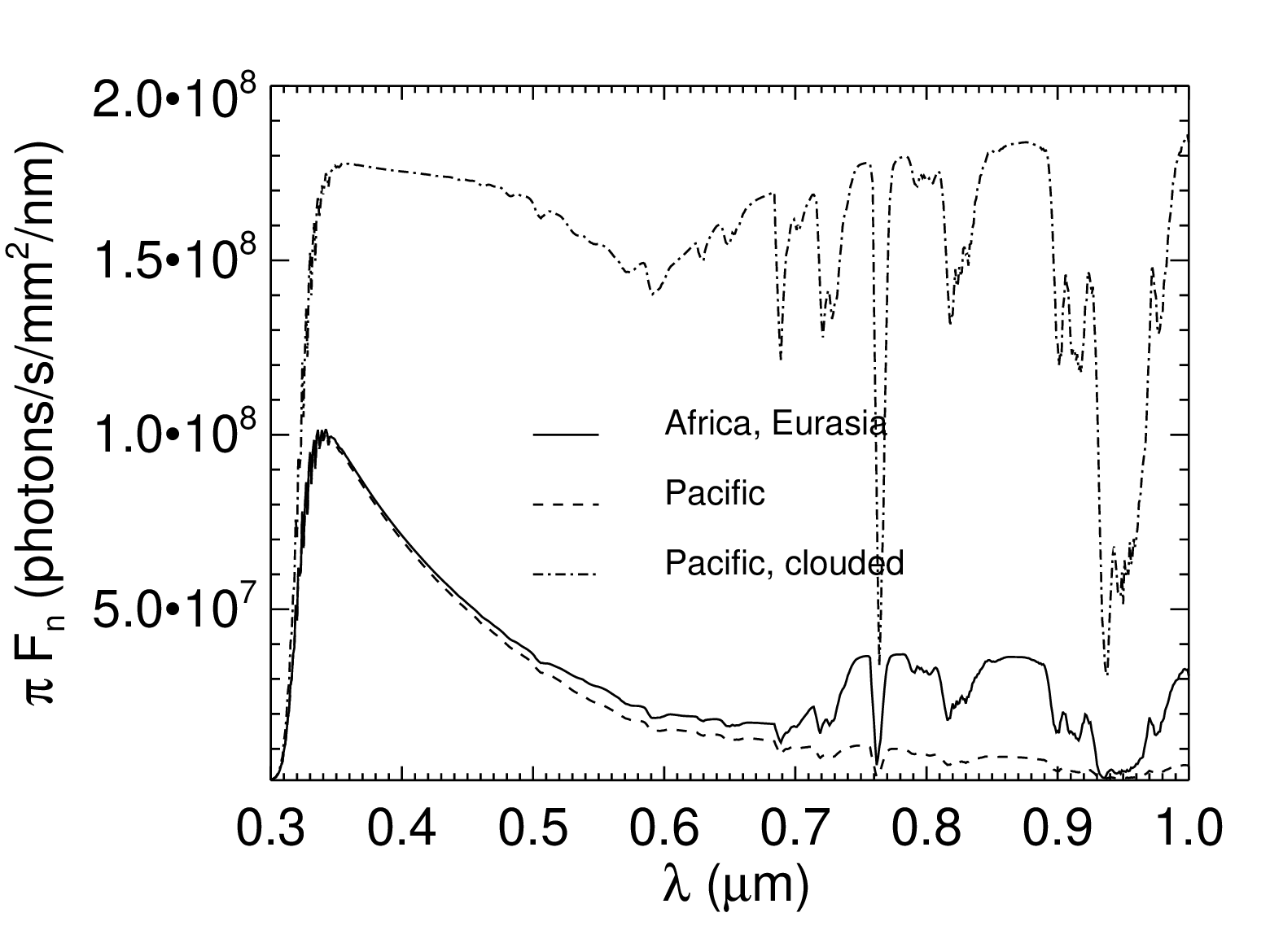}
\hspace{0.2cm}
\centering
\includegraphics[width=2.6in]{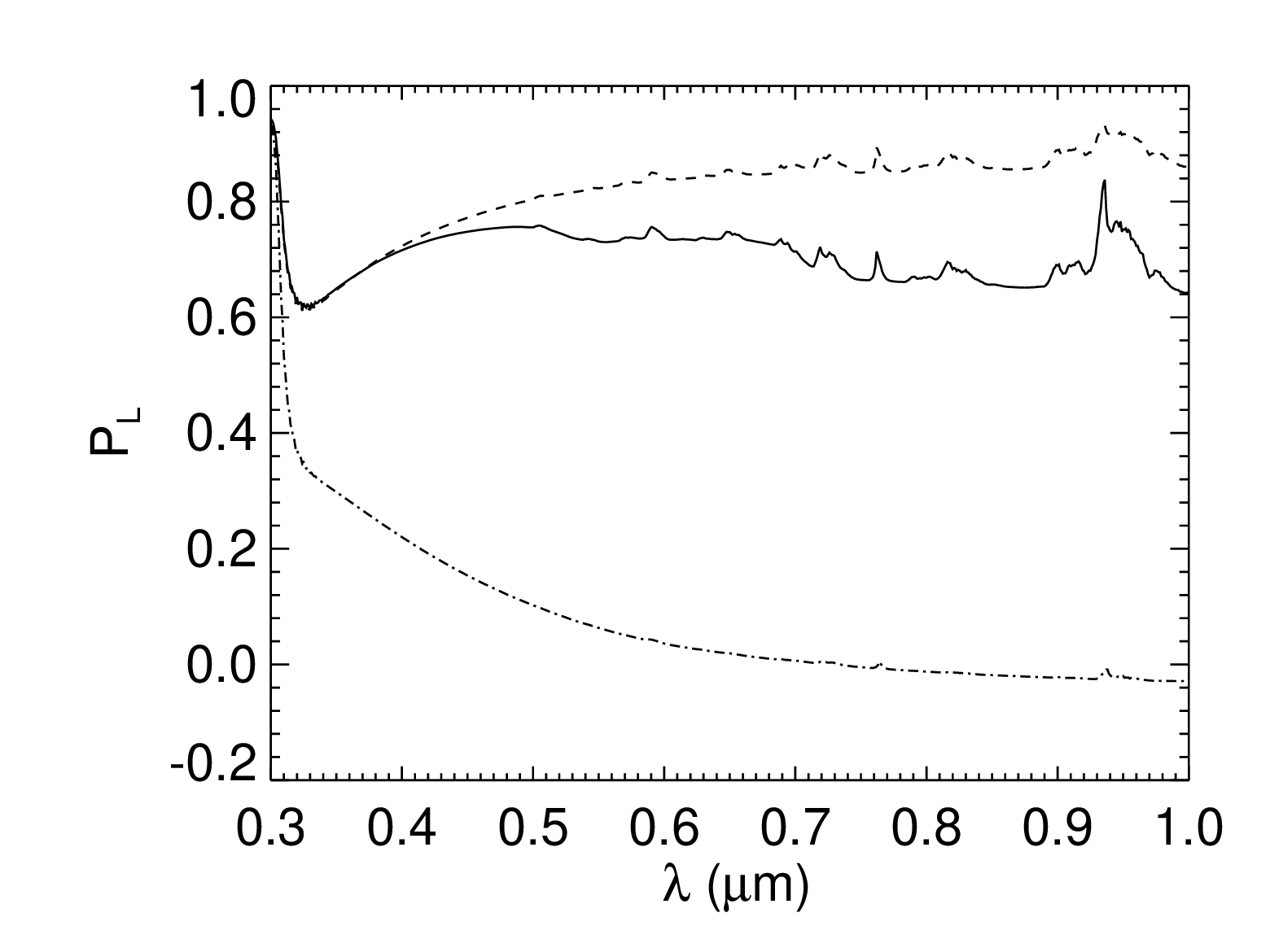}
\caption{Calculated flux $\pi F$ (left) and degree of linear
  polarisation $P_{\rm L}$ (right) of sunlight reflected by Earth as
  functions of $\lambda$, for $\alpha=90^\circ$: with Africa and
  Eurasia in view and no clouds (solid lines), with the Pacific ocean
  in view and no clouds (dashed lines) and when completely cloudy
  (dashed-dotted lines).}
\label{fig:model_fp}
\end{figure}

Figure~\ref{fig:model_fp} shows numerically calculated total fluxes
$\pi F$ and degrees of linear polarisation $P_{\rm L}$ as functions of
the wavelength $\lambda$.  The Earth's phase angle, $\alpha$, is
$90^\circ$ (from the moon, one would see a 'half' Earth).  The spectra
have been calculated using the radiative transfer algorithm described
in \citep[][]{stam08}, which assumes horizontally homogeneous model
planets. We used four model planets, covered by sand, forest, ocean,
or ice, combined with a cloud free or a completely cloudy atmosphere
(composed of the model A cloud particles of \citep{karalidi11}) with
an optical thickness of 10 (at 0.55~$\mu$m) and located between about
3~and~4~km.  The forest and ice surfaces are treated as Lambertian
reflectors, with albedos taken from the ASTER library. The ocean
surface is completely flat and black with a Fresnel reflecting
interface on top. The bi-directional and polarized reflection by the
sand surface is modeled using an optically thick ($\tau=20$ at all
$\lambda$) layer of dust particles \citep{laan09}, with a single
scattering albedo chosen such that the albedo agrees with that
measured from an airplane above the Sahara \citep{bierwirth09}.

To model the spectra of the horizontally {\em inhomogeneous} Earth, we
apply the weighted averages method \citep[][]{stam08} using the total
and polarized flux spectra of the horizontally homogeneous model
planets. In Fig.~\ref{fig:model_fp}, we have chosen the weighting
factors such that they represent a case in which Africa and Eurasia
are on the centre of the planetary disk and a case in which the
Pacific ocean is on the centre.  For comparison, the latter case is
also shown with a cloud layer.  For the solar flux that is incident on
the Earth, we assume a constant value throughout the whole spectrum
equal to the solar flux (in photons s$^{-1}$ mm$^{-2}$ nm$^{-1}$) as
measured by the GOME2 satellite instrument at 550~nm.

The flux and polarisation spectra of the cloud-free planets in
Fig.~\ref{fig:model_fp} clearly show the traces of the Earth's surface
and atmosphere. For a detailed explanation of the spectral features
due to gaseous absorption by O$_3$, O$_2$, and H$_2$O, see
\citep[][]{stam08}.  Longwards of 0.7~$\mu$m, the characteristic
red--edge albedo feature of the vegetation \citep{seager05} clearly
shows up in $\pi F$ when the continents are in full view: $\pi F$ is
higher by almost a factor of 5 (at 0.85~$\mu$m) than when the Pacific
is in view (a small fraction of this increase will be due to the sand
surface). The red--edge feature shows up as a decrease of $P_{\rm L}$
(of about 20\% in absolute value) because an increase in surface
albedo increases the amount of unpolarized light that is reflected
towards the observer.  Adding clouds to the model atmosphere increases
$\pi F$ strongly (except in the deepest gaseous absorption bands):
$\pi F$ is $\sim$12 times ($\sim$~23 times) higher at
$\lambda=$0.65~$\mu$m (0.85~$\mu$m). At the same time, the clouds
significantly decrease $P_{\rm L}$ at this phase angle
($\alpha=90^\circ$): $\sim$80\% at $\lambda=$0.65$\mu$m.  The model
planets used in Fig.~\ref{fig:model_fp} are either cloudfree of
completely cloudy. In reality, the Earth is only partly covered by
clouds (with a range of optical thicknesses) and the real flux and
polarisation spectra will be mixtures of the spectra that are shown
here.

\begin{figure}
\centering 
\includegraphics[width=2.6in]{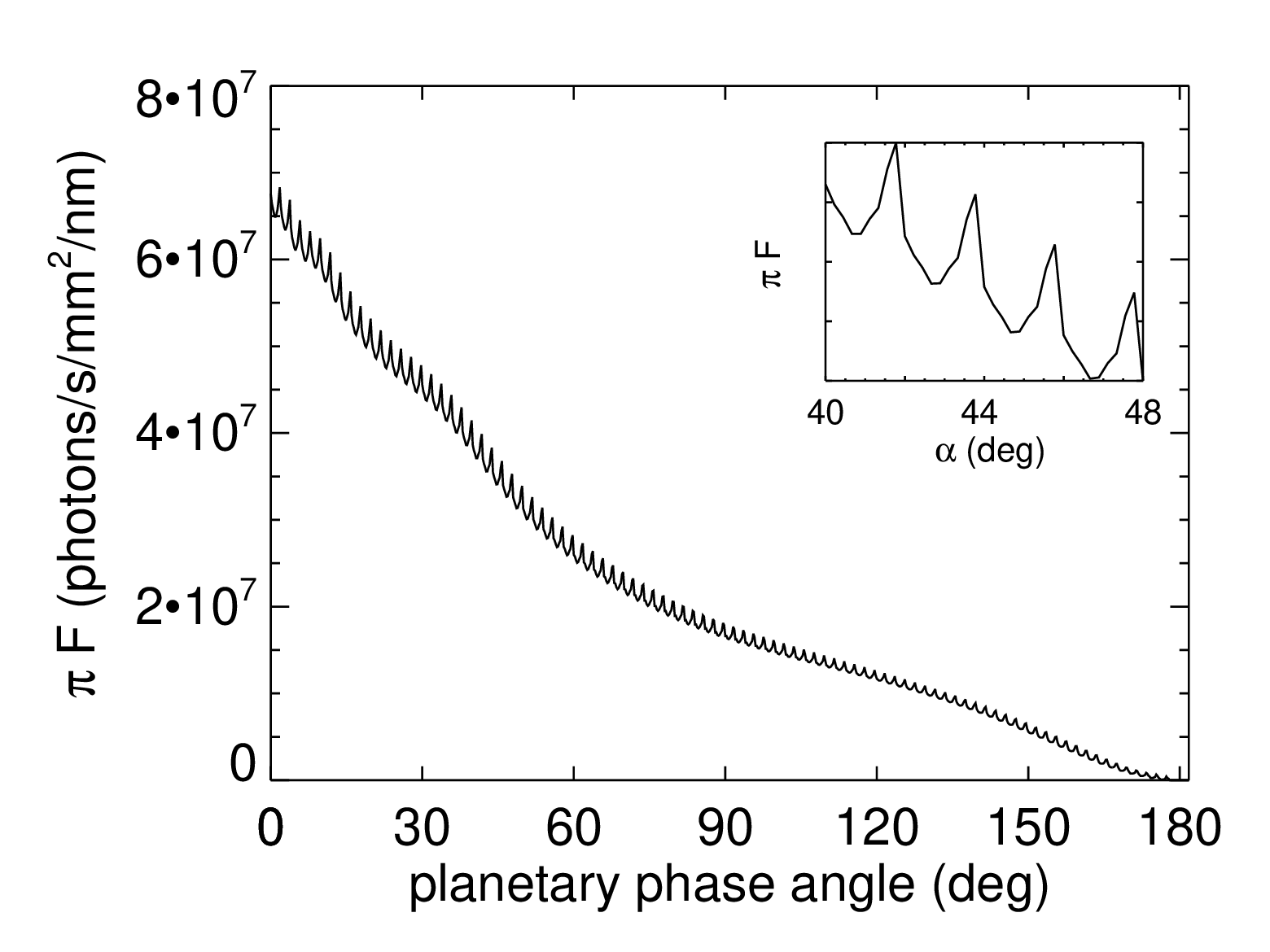}
\hspace{0.2cm}
\centering
\includegraphics[width=2.6in]{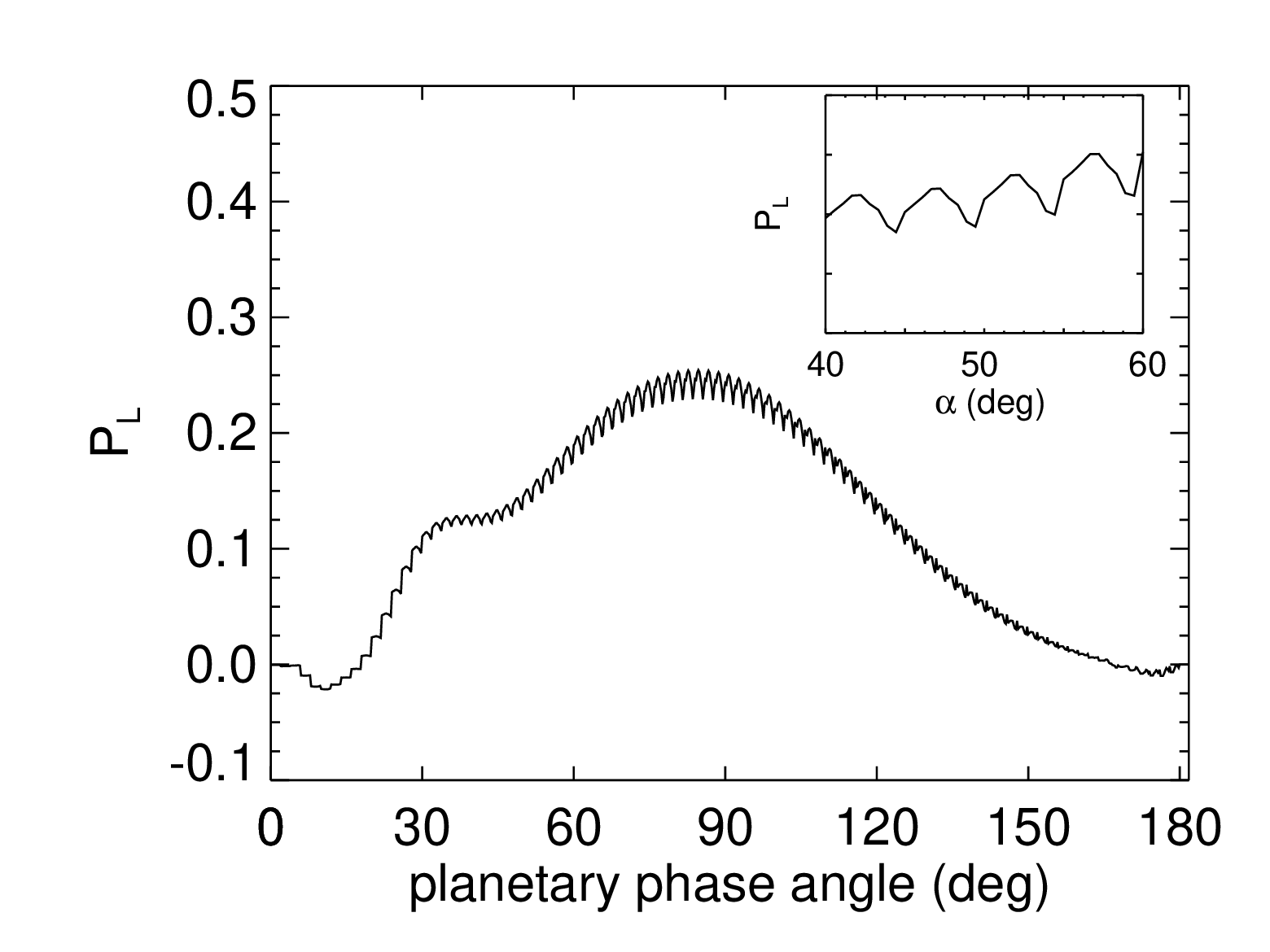}
\caption{Calculated $\pi F$ (left) and $P_{\rm L}$ (right) of sunlight
  reflected by a model Earth with 42\% cloud coverage as functions of
  $\alpha$, at 550 nm.}
\label{fig:cloudy_fp}
\end{figure}

In Fig.\ref{fig:cloudy_fp} we show $\pi F$ and $P_{\rm L}$ at
$\lambda$=550~nm, as functions of $\alpha$ for the model Earth that
has a cloud coverage of about 42~$\%$ (the cloud properties are the
same as in Fig.1).  The narrow features on top of the curves are due
to the daily rotation of the planet as it orbits its star, showing
ocean and/or continents through the holes in the clouds.  The
``bumps'' in the curves around $\alpha=30^\circ$ are due to the
primary rainbow: sunlight that has been scattered by the cloud
droplets once.  Clearly, the rainbow is much more pronounced in
$P_{\rm L}$ than in $\pi F$. Finding a rainbow in exoplanetary
polarisation signals will be a direct indication for the presence of
liquid water droplets in the planetary atmosphere.

%------------------------------------------------------------------------
\subsection{Circular Polarisation}

All known living material on Earth exhibits homo--chirality: sugars
and nucleic acids occur exclusively in the right--handed form, and
amino-acids and proteins in the left-handed form.  Homo--chirality
makes light scattered by organic material partially circularly
polarised, and circular polarimetric spectra of various samples of
biological material have been published
\citep[][]{wolstencroft04,sparks09,sterzik10}.  The reasons for
homo--chirality are unknown, but if similar evolutionary scenarios
naturally occur elsewhere in the universe, measuring $P_{\rm C}$ would
be a unique tool for the detection of life on exoplanets.  Since the
Earth is the only planet we know that has life on it, Earth
observations are the only way to empirically test this remote-sensing
method.  Some abiotic scattering processes (e.g. by optically active
atmospheric aerosols or minerals) may also give a measurable $P_{\rm
  C}$, but as shown in \citep[][]{sparks09} the wavelength dependence
of these signals is very different from that of the circular
polarisation of biological material \citep[][]{sparks09b}.

%------------------------------------------------------------------------
\section{The advantages of using the Moon as observation platform}
\label{Sec:Moon}

In order to build a comprehensive database of benchmark data of the
Earth as an exoplanet and to be able to fully test numerical
algorithms for signal simulation and planet characterisation, the
requirements on the flux and polarisation observations are as follows:

\noindent 1) Each observation of the Earth should be (nearly)
instantaneous, to observe different regions on the illuminated and
visible part of the Earth simultaneously and hence to capture the
effects of the differences in local solar zenith angles and viewing
angles.

\noindent 2) Observations should cover the Earth's diurnal cycle, to
capture the effects of different regions of the Earth emerging from
the night, and disappearing over the limb (or the other way around),
with the corresponding local changes in solar zenith and viewing
angles.

\noindent 3) The Earth should be observed at phase angles from
$\sim$0$^\circ$ ('full Earth') to $\sim$180$^\circ$ ('new Earth'),
with steps small enough to capture characteristic angular features in
the reflected $\pi F$ and $P$, such as the glint of sunlight reflected
by surface water and the rainbow of sunlight scattered in clouds.

\noindent 4) The observations should ideally cover all seasons to
capture the effects of changes in local solar zenith angles, polar
nights, weather and cloud patterns, and surface albedos.
      
Thanks to the monthly orbit of the Moon around the Earth and the tidal
locking of the Moon with respect to the Earth, a spectropolarimeter on
the lunar surface could observe the whole Earth, during each day, at
all phase angles (depending on the power source), and, in principle,
throughout the seasons.  Such whole Earth observations {\em cannot} be
obtained from (existing) artifical satellites, such as Low Earth Orbit
(LEO) remote-sensing satellites or geostationary satellites.

LEO satellites observe local regions on the Earth, and would require
several days to achieve global coverage.  In addition, a certain
location on Earth will always be observed under similar illumination
and viewing geometries (apart for seasonal variations of the local
solar zenith angle).  Currently, only the POLDER Earth-observing
satellite instrument has polarimetric capabilities (broadband, no
spectropolarimetry).  Geostationary satellites observe the same
hemisphere of the Earth all of the time. While these satellites do
capture the effects of the diurnal rotation and at the same time the
phase angle changes of the Earth, they cannot observe different
regions of the Earth, and their observations cannot teach us how to
derive a global distribution of oceans and continents from single
pixel measurements. There are currently no polarimeters onboard any
geostationary satellite.

Recent spectropolarimetric Earthshine observations
\citep[][]{sterzik12}, in which sunlight that has been reflected first
by the Earth and then by the moon is measured with Earth-based
instruments \citep[see e.g.][]{qiu03,sterzik09} confirm that
disk-integrated polarimetric observations are extremely sensitive to
the visible surface and atmosphere of the Earth.  At the same time,
discrepancies between theoretical predictions and observations
demonstrate that multi-epoch observations of the Earth are needed to
constrain the models.  The major drawback of Earthshine observations
is that the properties of the lunar surface are not known well,
especially when polarisation is involved. This makes the modelling
enormously more difficult than in the case of observations from space
(including the Moon).  Ground-based Earthshine observations are also
hampered by background contamination from the sunlit fraction of the
Moon, and do not allow the same phase angle coverage (both in range
and in angular resolution) and are unable to capture the full diurnal
rotation.

Finally, a number of non--dedicated missions (e.g. Voyager 1, and more
recently Deep Impact) have taken snapshots of the Earth.\footnote{for
  a nice overview see:
  http://planetary.org/explore/topics/earth/spacecraft.html} These
observations, while often providing interesting data, do not cover the
diurnal rotation nor the phase angle range nor the seasonal
effects. There have been no polarimetric observations performed by
such missions.

A spectropolarimeter could be put onboard a specially designed
satellite in an orbit that allows performing the required
observations. That orbit would, however, probably closely resemble the
orbit of the moon. Including the instrument on a Lunar Lander thus
seems a straightfoward and economical choice.

%------------------------------------------------------------------------
\section{LOUPE: the Lunar Observer for Unresolved Polarimetry of Earth}
\label{Sec:LOUPE}

\begin{figure}[t]
\centering
\vspace*{-0.5cm}
\includegraphics[width=\textwidth]{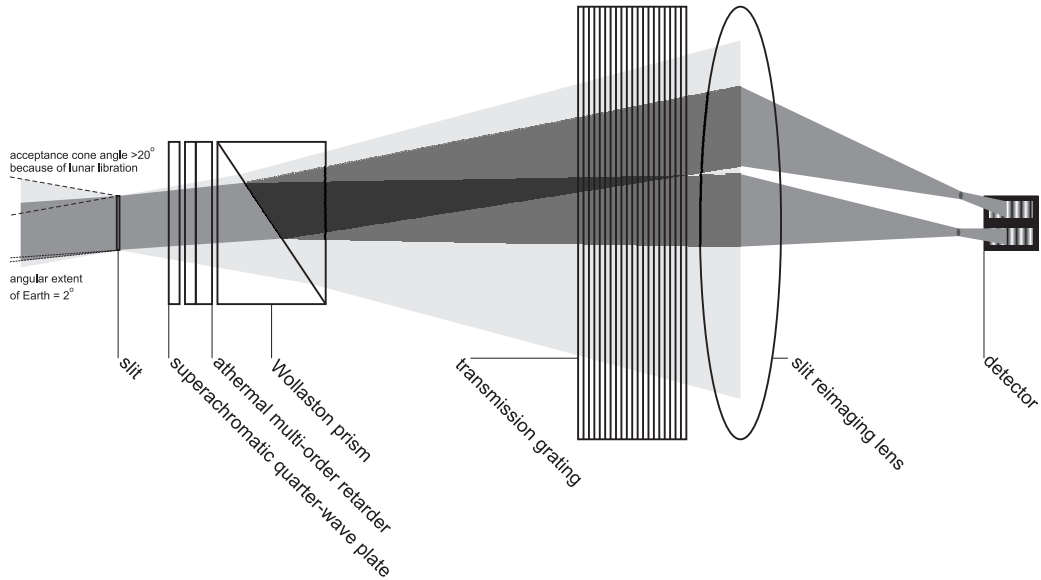}
\caption{Schematic depiction of LOUPE option 1A (only linear
  spectropolarimetry, no spatial resolution) The scale is
  approximately 1:1. Such an instrument only needs to be roughly
  pointed towards the Earth as the instruments accepts light from all
  angles within the range determined by the lunar libration.}
\label{LOUPE1A}
\end{figure}

The Lunar Observatory for Unresolved Polarimetry of Earth (LOUPE)
shall fulfill the following requirements:
\begin{itemize}
\item It performs spectropolarimetric observations of the light from
  the Earth's disk (at least) at visible wavelengths (400--800 nm).
\item The spectral resolution for the polarimetry shall be $\sim$20
  nm, while the O$_2$A band ($\sim$0.76~$\mu$m) is resolved in the
  flux spectrum. Limited spectropolarimetry can be performed within
  this and other bands.
\item Data is collected on an hourly basis to resolve the Earth's
  rotation, and span at least a month to cover a full range of phase
  angles.
\item The instrument is small and robust.
\end{itemize}
For the polarimetry, we explore two different scenarios:

\noindent 
1) \textit{Linear spectropolarimetry only}. For this we adopt the
spectral modulation approach \citep{snik09}. Using a combination of
standard solid-state polarization optics (see Fig.~\ref{LOUPE1A}), the
total flux spectrum is multiplied by a sinusoidal modulation for which
the amplitude scales with $P_{\rm L}$, and the phase is determined by
the angle of polarization.  This novel polarimetric concept is being
applied in the SPEX instruments. The SPEX prototype exhibits excellent
polarimetric performance \citep{vanharten11}.

\noindent 
2) \textit{Linear {\em and} circular spectropolarimetry}. This
implementation is more challenging as the data dimensionality is
larger, and, moreover, $P_{\rm C}$ ($\sim$10$^{-4}$) is much smaller
than the average $P_{\rm L}$ ($\sim$10$^{-4}$ versus $\sim 0.1$).  The
spectral modulation approach in \citep{okakato99} yields three
modulation periods that contain information on the complete flux
vector. The modulation approach introduced by \citep{sparks12} yields
similar modulations, but along the slit direction.

Various options can be identified for spatial resolution and pointing:

\noindent
A) The instrument itself averages the light from the Earth's disk
($\sim$2$^\circ$ diameter). Because the disk is surrounded by black
space, this requires only course pointing.  The acceptance angle of
the instrument should be wide enough to take lunar libration ($\pm
8^\circ$) into account.

\noindent
B) The instrument spatially resolves the Earth's disk to obtain data
of e.g.~just the Amazonian rainforest to maximize the circular
polarization signal. Such spatial information can be attained by using
a scanning slit or an integral field unit. In any case, accurate
pointing and potentially scanning should be implemented. Averaging
over the Earth's disk is then performed in the data pipeline.

A sketch of the most basic implementation (1A: only linear
spectropolarimetry and no spatial resolution) of LOUPE is presented in
Fig.~\ref{LOUPE1A}.

%------------------------------------------------------------------------
\section{Summary and conclusions}
\label{Sec:summary}

We present LOUPE, the Lunar Observatory for Unresolved Polarimetry of
Earth. LOUPE is a small and robust spectropolarimeter that can observe
the Earth as if it were an exoplanet from a vantage point on the lunar
surface.  The Moon has a unique position with respect to Earth and can
provide us with a unique platform from where we can observe the Earth
as an exoplanet.  From the Moon, LOUPE will be able to observe the
whole disk of the Earth, all of the time, at most phase angles and
throughout the year.

LOUPE measures the total flux and state (degree and direction) of
polarisation of sunlight that is reflected by the Earth.  Polarimetry
appears to be a strong tool for the characterisation of exoplanets,
allowing the retrieval of the composition and structure of a planet's
atmosphere and surface (if present). In particular, the degree of
linear polarisation can give us information on the presence of liquid
water clouds and the degree of circular polarisation on the presence
of life.  LOUPE measurements would be used as a benchmark for future
Earth-like exoplanet observations and to test numerical algorithms for
the retrieval of planet properties from such observations.

%% not want to use model1c-num-names.bst.

%\bibliographystyle{elsarticle-harv}
%\bibliography{references}

\begin{thebibliography}{25}
\expandafter\ifx\csname natexlab\endcsname\relax\def\natexlab#1{#1}\fi
\expandafter\ifx\csname url\endcsname\relax
  \def\url#1{\texttt{#1}}\fi
\expandafter\ifx\csname urlprefix\endcsname\relax\def\urlprefix{URL }\fi

%------------------------------------------------------------------------
\bibitem{beaulieu10}
{Beaulieu}, J.~P., {Kipping}, D.~M., {Batista}, V., {Tinetti}, G., {Ribas}, I.,
{Carey}, S., {Noriega-Crespo}, J.~A., {Griffith}, C.~A., {Campanella}, G.,
{Dong}, S., {Tennyson}, J., {Barber}, R.~J., {Deroo}, P., {Fossey}, S.~J.,
{Liang}, D., {Swain}, M.~R., {Yung}, Y., {Allard}, N., 2010. 
{Water in the atmosphere of HD 209458b from 3.6-8 {$\mu$}m IRAC photometric
observations in primary transit}. 
Monthly Notices of the Royal Astron. Soc. 409, 963--974.

\bibitem{bierwirth09}
{Bierwirth}, E., {Wendisch}, M., {Ehrlich}, A., {Heese}, B., {Tesche}, M.,
{Althausen}, D., {Schladitz}, A., {M{\"u}ller}, D., {Otto}, S., {Trautmann},
T., {Dinter}, T., {von Hoyningen-Huene}, W., {Kahn}, R., 2009. 
{Spectral surface albedo over Morocco and its impact on radiative forcing of 
Saharan dust}. Tellus Series B: Chemical and Physical Meteorology B 61, 252--269.

\bibitem{cassan12}
{Cassan}, A., {Kubas}, D., {Beaulieu}, J.-P., {Dominik}, M., {Horne}, K.,
{Greenhill}, J., {Wambsganss}, J., {Menzies}, J., {Williams}, A.,
{J{\o}rgensen}, U.~G., {Udalski}, A., {Bennett}, D.~P., {Albrow}, M.~D.,
{Batista}, V., {Brillant}, S., {Caldwell}, J.~A.~R., {Cole}, A., {Coutures},
C., {Cook}, K.~H., {Dieters}, S., {Prester}, D.~D., {Donatowicz}, J.,
{Fouqu{\'e}}, P., {Hill}, K., {Kains}, N., {Kane}, S., {Marquette}, J.-B.,
{Martin}, R., {Pollard}, K.~R., {Sahu}, K.~C., {Vinter}, C., {Warren}, D.,
{Watson}, B., {Zub}, M., {Sumi}, T., {Szyma{\'n}ski}, M.~K., {Kubiak}, M.,
{Poleski}, R., {Soszynski}, I., {Ulaczyk}, K., {Pietrzy{\'n}ski}, G.,
{Wyrzykowski}, {\L}., 2012. 
{One or more bound planets per Milky Way star from microlensing observations}. 
Nature 481, 167--169.

\bibitem{ford01}
{Ford}, E.~B., {Seager}, S., {Turner}, E.~L., 2001. 
{Characterization of extrasolar terrestrial planets from diurnal photometric 
variability}. Nature 412, 885--887.

\bibitem{hansenhovenier74}
{Hansen}, J.~E., {Hovenier}, J.~W., 1974. 
{Interpretation of the polarization of Venus.} 
J. Atmos. Sci. 31, 1137--1160.

\bibitem{hansentravis74}
{Hansen}, J.~E., {Travis}, L.~D., 1974. 
{Light scattering in planetary atmospheres.} 
Space Sci. Revs. 16, 527--610.

\bibitem{hovenier04}
{Hovenier}, J.~W., {Van der Mee}, C., {Domke}, H., 2004. 
{Transfer of polarized light in planetary atmospheres: basic concepts and practical 
methods}. Vol. 318 of Astrophysics and Space Science Library.

\bibitem{kaltenegger09}
{Kaltenegger}, L., {Traub}, W.~A., 2009. 
{Transits of Earth-like Planets}.
Astrophys. J. 698, 519--527.

\bibitem{karalidi11b}
{Karalidi}, T., {Stam}, D.~M., 2012 (in preparation).
{Modelling horizontally inhomogeneous exoplanets in flux and polarisation}.

\bibitem{karalidi11}
{Karalidi}, T., {Stam}, D.~M., {Hovenier}, J.~W., 2011. 
{Flux and polarisation spectra of water clouds on exoplanets}. 
Astron. Astrophys. 530, A69+.

\bibitem{keller10}
{Keller}, C.~U., {Schmid}, H.~M., {Venema}, L.~B., {Hanenburg}, H., {Jager},
R., {Kasper}, M., {Martinez}, P., {Rigal}, F., {Rodenhuis}, M., {Roelfsema},
R., {Snik}, F., {Verinaud}, C., {Yaitskova}, N., 2010. 
{EPOL: the exoplanet polarimeter for EPICS at the E-ELT}. 
In: Society of Photo-Optical Instrumentation Engineers (SPIE) Conference Series,
Vol. 7735.

\bibitem{laan09}
{Laan}, E.~C., {Volten}, H., {Stam}, D.~M., {Mu{\~n}oz}, O., {Hovenier}, J.~W.,
{Roush}, T.~L., 2009. 
{Scattering matrices and expansion coefficients of martian analogue palagonite particles}. 
Icarus 199, 219--230.

\bibitem{mayorqueloz95}
{Mayor}, M., {Queloz}, D., 1995. 
{A Jupiter-Mass Companion to a Solar-Type Star}. 
Nature 378, 355--+.

\bibitem{millerricci10}
{Miller-Ricci}, E., {Fortney}, J.~J., 2010. 
{The Nature of the Atmosphere of the Transiting Super-Earth GJ 1214b}. 
Astrophys. J. Lett. 716, L74--L79.

\bibitem{mishchenko97}
{Mishchenko}, M.~I., {Travis}, L.~D., 1997. 
{Satellite retrieval of aerosol properties over the ocean using polarization as well 
as intensity of reflected sunlight}. 
J. Geophys. Res. 102, 16989--17014.

\bibitem{okakato99} 
Oka, K., \& Kato, T. 1999, 
{Spectroscopic polarimetry with a channeled spectrum}.
Optics Lett., 24, 1475--1477.

\bibitem{qiu03}
{Qiu}, J., {Goode}, P.~R., {Pall{\'e}}, E., {Yurchyshyn}, V., {Hickey}, J.,
{Monta{\~n}{\'e}s Rodriguez}, P., {Chu}, M.-C., {Kolbe}, E., {Brown}, C.~T.,
{Koonin}, S.~E., 2003. 
{Earthshine and the Earth's albedo: 1. Earthshine observations and measurements 
of the lunar phase function for accurate measurements of the Earth's Bond albedo}. 
J. Geophys. Res. (Atmos) 108, 4709--.

\bibitem{seager05}
{Seager}, S., {Turner}, E.~L., {Schafer}, J., {Ford}, E.~B., 2005.
{Vegetation's Red Edge: A Possible Spectroscopic Biosignature of
Extraterrestrial Plants}. Astrobiology 5, 372--390.

\bibitem{seager00}
{Seager}, S., {Whitney}, B.~A., {Sasselov}, D.~D., 2000. 
{Photometric Light Curves and Polarization of Close-in Extrasolar Giant Planets}. 
Astrophys. J. 540, 504--520.

\bibitem{snik09} 
Snik, F., Karalidi, T., \& Keller, C. U. 2009, 
{Spectral Modulation for full linear polarimetry}
Appl. Optics, 48, 1337--.

\bibitem{sparks09}
{Sparks}, W.~B., {Hough}, J., {Germer}, T.~A., {Chen}, F., {Dassarma}, S.,
{Dassarma}, P., {Robb}, F.~T., {Manset}, N., {Kolokolova}, L., {Reid}, N.,
{Macchetto}, F.~D., {Martin}, W., 2009. 
{Detection of circular polarization in light scattered from photosynthetic microbes}. 
Proc. National Academy of Science 106, 7816--7821.

\bibitem{sparks09b}
{Sparks}, W.~B., {Hough}, J., {Kolokolova}, L., {Germer}, T.~A., {Chen}, F.,
{DasSarma}, S., {DasSarma}, P., {Robb}, F.~T., {Manset}, N., {Reid}, I.~N.,
{Macchetto}, F.~D. , {Martin}, W., 2009. 
{Circular polarization in scattered light as a possible biomarker}. 
JQSRT 110, 1771--1779.


\bibitem{sparks12} 
Sparks, W.~B., Germer, T.~A., MacKenty, J., \& Snik, F. 2012, 
Appl. Optics (to be submitted).

\bibitem{stam08}
{Stam}, D.~M., 2008. {Spectropolarimetric signatures of Earth-like
extrasolar planets}. Astron. Astrophys. 482, 989--1007.

\bibitem{sterzik10}
{Sterzik}, M., {Bagnulo}, S., {Azua}, A., {Salinas}, F., {Alfaro}, J.,
{Vicuna}, R., 2010. 
{Astronomy Meets Biology: EFOSC2 and the Chirality of Life}. 
The Messenger 142, 25--27.

\bibitem{sterzik12}
{Sterzik}, M., {Bagnulo}, S., {Pall\'e}, E., 2012. 
{Biosignatures as revealed by spectropolarimetry of Earthshine}.
Nature  483, 64-66.

\bibitem{sterzik09}
{Sterzik}, M.~F., {Bagnulo}, S., 2009. 
{Search For Chiral Signatures in the Earthshine}. 
In: {K.~J.~Meech, J.~V.~Keane, M.~J.~Mumma, J.~L.~Siefert,
\& D.~J.~Werthimer } (Ed.), Bioastronomy 2007: Molecules, Microbes and
Extraterrestrial Life. Vol. 420 of Astronomical Society of the Pacific
Conference Series, 371--.

\bibitem{tinetti06c}
{Tinetti}, G., {Meadows}, V.~S., {Crisp}, D., {Fong}, W., {Fishbein}, E.,
{Turnbull}, M., {Bibring}, J.-P., 2006. 
{Detectability of Planetary Characteristics in Disk-Averaged Spectra. I: The Earth Model}. 
Astrobiology 6, 34--47.

\bibitem{vanharten11} 
van Harten, G., Snik, F., Rietjens, J.H.H, Smit, J.M., de Boer, J., Diamantopoulou, R.,
Hasekamp, O.P., Stam, D.M., Keller, C.U. Laan, E.C., Verlaan, A.L., Vliegenthart, W.A.,
ter Horst, R., Navarro, R., Wielinga, K., Hannemann, S., Moon, S.G., and Voors, R., 2011.
Prototyping for the Spectropolarimeter for Planetary EXploration (SPEX): 
calibration and sky measurements, Proc. SPIE 8160, 81600Z.

\bibitem{wordsworth11}
{Wordsworth}, R.~D., {Forget}, F., {Selsis}, F., {Millour}, E., {Charnay}, B.,
{Madeleine}, J.-B., 2011. 
{Gliese 581d is the First Discovered Terrestrial-mass Exoplanet in the Habitable Zone}. 
Astrophys. J. Lett. 733, L48.


\bibitem{wolstencroft04}
{{Wolstencroft}, R.~D.} 2004.
{errestrial and Astronomical Sources of Circular Polarisation: A Fresh Look at the Origin of OF Homochirality on Earth}.
Bioastronomy 2002: Life Among the Stars, 213


\end{thebibliography}


\end{document}